\documentclass[aps,prb,preprint,12pt]{revtex4-1}
\usepackage{amssymb,euscript,latexsym,amsmath}
\usepackage{graphicx}
\usepackage{tikz,pgfplots}
\usepackage{amsmath}
\usepackage{tikz}
\usepackage{hyperref}
\usepackage[makeroom]{cancel}
\begin{document}
\title{The chiral knife edge: a simplified rattleback to illustrate spin inversion}
\author{Eduardo A. Jagla}
\affiliation{Comisi\'on Nacional de Energ\'ia At\'omica and Instituto Balseiro\\ CNEA, CONICET, UNCUYO\\
Av. E. Bustillo 9500 S. C. de Bariloche, Argentina}
\author {Alberto G.  Rojo }
\affiliation{Department of Physics, Oakland University, Rochester, MI 48309}

\begin{abstract}
    We present the chiral knife edge  rattleback, an alternative version of previously presented systems that exhibit spin inversion.  We offer a full treatment of the model using qualitative arguments, analytical solutions  as well as numerical results. We treat   a reduced, one--mode problem which not only contains the essence of the physics of spin inversion, but that also exhibits an unexpected connection  to the Chaplygin sleigh, providing new insight into the non-holonomic structure of the problem.  We also present exact results for the full problem together with estimates of the time between inversions that agree with previous results in the literature.
\end{abstract}

\maketitle

\section{Introduction}

The rattleback, or {\em celt}, is a  boat-shaped stone (commercially available as a 
toy)  which exhibits unintuitive behavior that at first glance seems to defy the law of conservation of angular momentum. In its usual version it consists of a simple rigid body with a  semi-ellipsoidal bottom and an uneven distribution of mass so that the axes of symmetry of the ellipsoid do not coincide with the principal axes of inertia of the rigid body. 

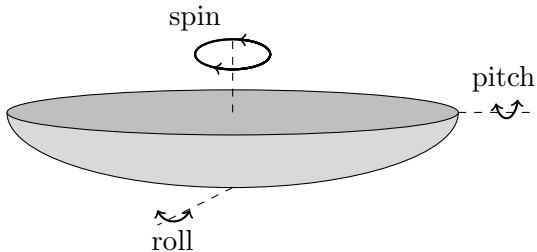
\begin{figure}[h]
\centering
\begin{tikzpicture}
     \filldraw[fill=gray!30]
(0,0)  arc(-180:0:3cm and 1cm);
   \filldraw[fill=gray!50] (6cm,0)  arc(0:360:3cm and .3cm);
   \draw [dashed](3cm,0cm)--(3cm,1cm);
   \draw[thick,<-] (2.75cm,.6cm) arc (-120:440:.5cm and .2 cm);
    \draw[thick,->] (2.75cm,.6cm) arc (-120:440:.5cm and .2 cm);
   \node at (2.5cm,1.25cm) {spin};
      \draw[thick,<->] (6.5cm,.12cm) arc (-160:-10:.15cm and .3 cm);
        \draw [dashed](6cm,0cm)--(7cm,0cm);
           \node at (6.6cm,.45cm) {pitch};
  \draw [dashed](3cm,-1cm)--(2cm,-1.5cm); 
        \draw[thick,<->] (2.cm,-1.3cm) arc (210:330:.25cm and .3 cm);    
              \node at (2.2cm,-1.7cm){roll};    
\end{tikzpicture}
\caption{Scheme of a traditional boat-shaped rattleback. The two oscillation and the rotation degrees of freedom are indicated.}
\label{PitchRoll}
\end{figure}

As with a symmetric semi-ellipsoidal  top, the rattleback can oscillate with respect to two horizontal axes in modes usually called ``rolling" and ``pitching" as shown schematically in Figure \ref{PitchRoll}.
 When the rattleback  is spun in one direction,
 it quickly starts to roll up and down while the rotation velocity first decreases, and eventually changes sign.
After a few turns   it soon begins pitching until the rotation change direction again. 
Were it not for unavoidable mechanical loses this periodic inversion of the rotation direction would continue indefinitely. 
The misalignment between the principal axes of inertia of the body and those
of the curvature at the contact point (generating a definite {\em chirality} in the system) couples the spinning motion with the
pitching and rolling oscillations. 
As a result of this misalignment, the frictional contact force  creates a torque
about the center of mass causing the top
to invert its motion. 

\begin{figure}[h]
\centering
\includegraphics[width=6cm,clip=true]{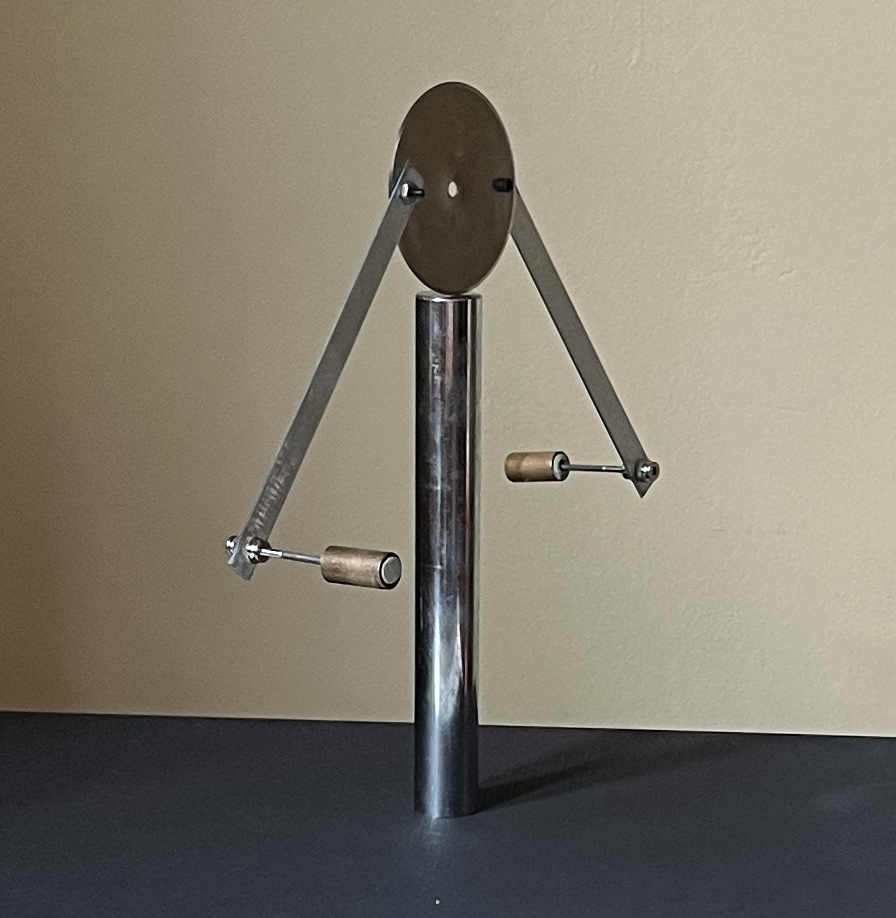}
\caption{Chiral rattleback built using a pizza cutter disk as a knife edge. The cylinder is for support only, and is kept fixed.
}
\label{Pizza}
\end{figure}

Most of previous analyses of the rattleback \cite{Walker1,Bondi,Casea,Walker2} describe the  motion by approximating the contact surface of 
the body as an ellipsoid whose principal axes are rotated with respect to the principal axes of inertia. The equations of motion 
are then treated in various approximations and solved numerically to illustrate the spin inversion. In this paper we 
propose an alternative version of the rattleback on the form of a knife edge with hanging masses that can execute the 
rolling and pitching motion, with the center of mass below the point of contact.  We performed some experimental tries with the physical model shown in Figure \ref{Pizza}

Our derivation of the equations  of motion is simpler than that of previous treatments, and these equations are even suitable for pedagogical expositions that illuminate the origin of spin inversion.
In order to distill the essence of 
the mechanism of spin inversion we first treat a simplified model in which we freeze one of the oscillating 
modes--the single mode rattleback. 

In section \ref{Singlemode} we describe the simplified single mode case, present a qualitative explanation of its behavior and solve the full non-holonomic equations of motion using a few reasonable approximations. The resulting equations are simple and amenable to a transparent interpretation of the origin of the spin inversion. In addition, we rewrite the equations in terms of the amplitudes of motion of the oscillating and spinning mode, and retrieve by -in our opinion- a more direct and ``microscopic" route the equations of motion (restricted to one mode) proposed by Tokieda and collaborators\cite{Moffatt,Yoshida}. In addition we find a puzzling and unexpected equivalence of our single mode rattleback with the Chaplygin sleigh, one of the classic irreversible non-holonomic systems. 

In section \ref{Twomodes} we extend the treatment to the knife edge rattleback with two modes. We  follow the standard non-holonomic program to obtain analytical expressions  for the equations of motion. We also present numerical solutions that show an equivalence with the traditional treatments.


\section{Single mode knife edge rattleback}
\label{Singlemode}

\begin{figure}[h]
\centering
\includegraphics[scale=0.4]{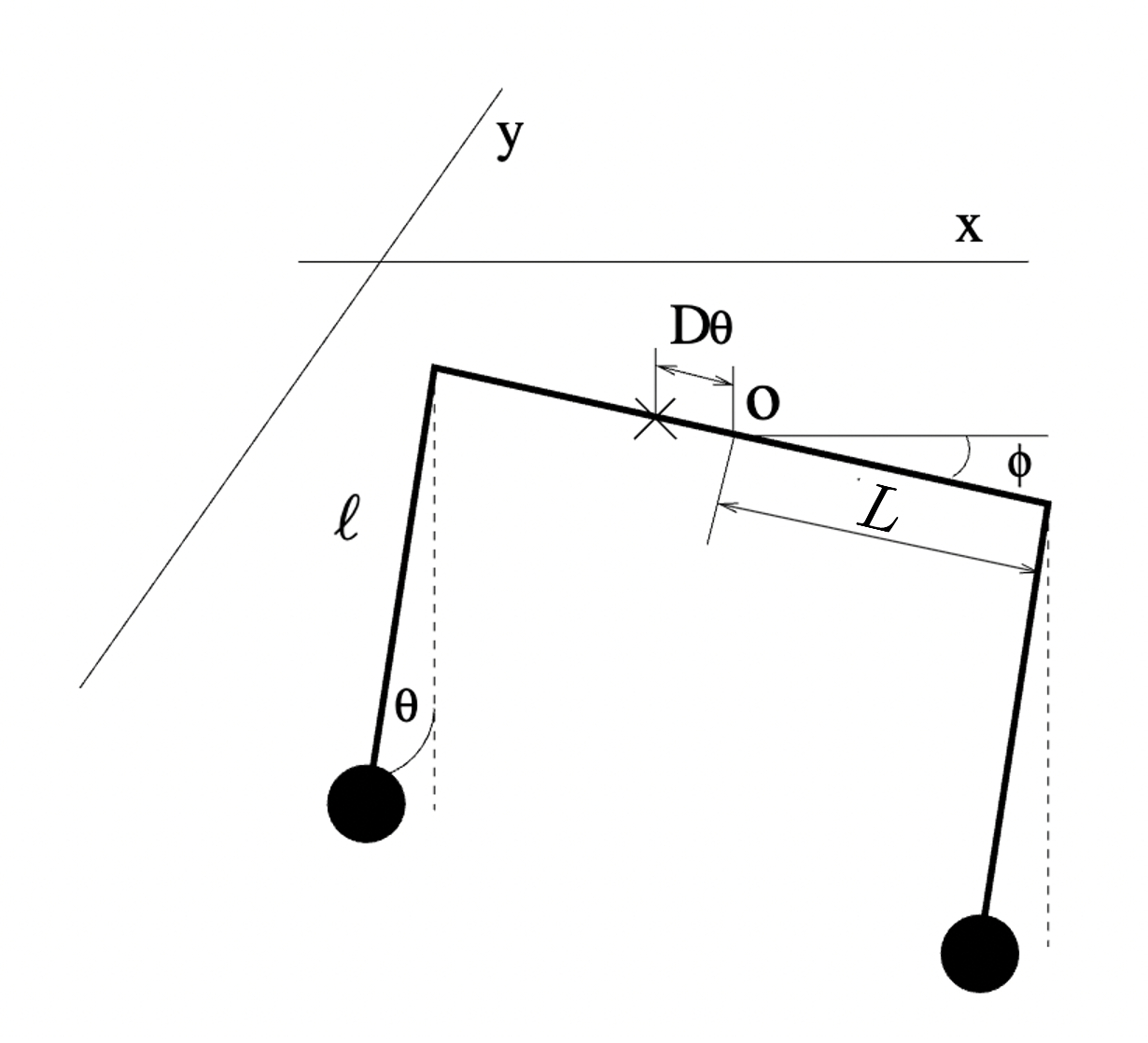}
\caption{Sketch of the single mode rattleback. The horizontal bar lies completely on the $x,y$ plane, at an angle $\phi$ with respect to $\hat x$. The instantaneous contact point (indicated by $\times$ and having zero velocity) is at distance $D\theta$ from the center point $O$.
}
\label{figg1}
\end{figure}

The rattleback effect is driven by a shift of the 
supporting point with the oscillation angles.
We first attempt to the simplest description of the phenomenon. 
We consider a single oscillating mode, that plays the role of either pitching or rolling in the usual nomenclature of the rattleback, coupled to a spin mode, and assume that the second oscillating mode is ``frozen".
Specifically, our model consists of a rigid  mass-less bar --a knife edge-- of length $2L$ that stays and moves on the $(x,y)$ plane. Two mass-less segments of length $\ell$ are attached to the ends of the bar. These two segments have masses of magnitude $M/2$ attached to their free ends. The set of the three mass-less segments and the two point masses form a rigid body.
The system can spin with angular velocity $\dot \phi$ around the $z$ axis and execute small oscillations of angle $\theta$ around the vertical, as illustrated in  Figure \ref{figg1}. The choice of this particular geometry  is informed by experiments we did with a chiral knife edge that, in order to be stable, requires the center of mass to be below the point of support.  

The configuration space of the system is determined by the two angles $\theta$ and $\phi$, and the position $\mathbf x$
of the center of the horizontal bar (point $O$ in Fig. \ref{figg1}) on the $x$,$y$ plane.
To completely define the dynamics of the model we need to specify the sliding conditions of the horizontal bar on the plane. We will 
assume that at any moment there is a single contact point $\mathbf x_C$ with a non-slip condition between the bar and the plane. The 
instantaneous zero velocity of the physical contact point is the additional ingredient that completely defines the dynamics. 
Yet, the non-trivial condition from which the subtle properties of the system will emerge is the change of contact point with the value of $\theta$. 
We impose that the instantaneous contact point is at a distance $D\theta$ from the center  $\mathbf x$  of the bar:
$$
\mathbf x_C=\mathbf x + D\theta \hat{\mathbf {u}}_\phi
$$
where  $ \hat{\mathbf {u}}_\phi$  a unit vector in the direction of the bar.
This prescription, that introduces chirality into the system, will be fully justified within the more complete 
modeling that includes the two oscillating modes of the system and a more realistic knife edge geometry, to be presented in section
\ref{Twomodes}.

\subsection{Qualitative explanation of the spin inversion}
\label{Qualitative}
\begin{figure}[h]
\begin{center}
\includegraphics[width=6cm,clip=true]{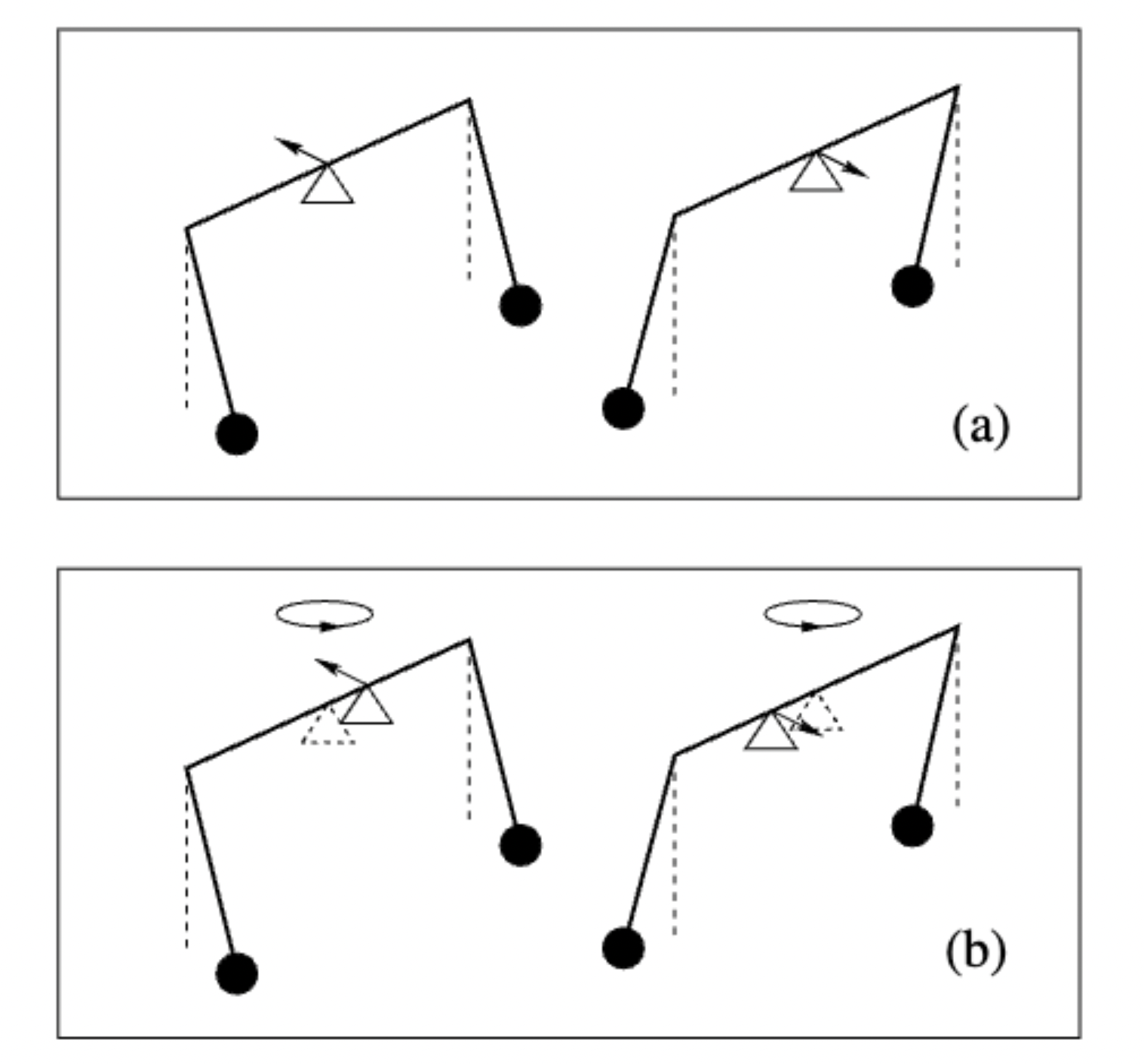}
\end{center}
\caption{Origin of the oscillation-spin coupling in the single mode rattleback. In (a) the supporting point is kept fixed, and the reaction force due to the oscillation (indicated with the arrows) does not generate any torque onto the system. In (b) the supporting point shifts 
proportionally to $\theta$, and therefore the reaction force produces a torque always with the same orientation.
}
\label{f2}
\end{figure}

Consider the case in which $D=0$, namely, the contact point is always the middle point of the bar. Also, let us take $\dot\phi=0$.
Under these conditions all we have is a simple pendulum which, for  small values of $\theta$,  executes a harmonic motion of the form
\begin{equation}
    \theta(t) \propto \cos \omega_0 t.
\end{equation}
 This oscillation does not couple to $\phi$. 
 During the oscillation, there is a friction force (provided by the constraint) of the form
\begin{equation}
f(t)\propto  \ddot\theta(t)\propto  -\theta(t)
\end{equation}
acting at the pendulum's support point, as sketched in Fig.\ref{f2}(a).

If the contact point changes with $\theta$ (i.e., $D\ne0$), the force $f(t)$ 
will now be applied 
at a distance $D\theta$ away from the center $O$ of the bar (Fig. \ref{f2}(b)). This force now generates a torque with respect to $O$, giving rise to an acceleration of $\phi$, in the form

\begin{equation}
\ddot \phi\sim f \times D\theta \sim -D\theta ^2 
\end{equation}
The torque appears in a well defined 
direction, independently of the sign of $\theta$, and is proportional to the energy of the oscillatory mode. Note that this fact appeared as a postulate in 
one of the first full mathematical treatments of the rattleback\cite{Hubbard2}.

This pedagogical exposition of the oscillation-rotation coupling is at the heart of the rattleback effect in more complex set ups.
Now we proceed  to the full analysis of this single mode rattleback. 

\subsection{Full analysis}

The unconstrained Lagrangian of the system in Fig. \ref{figg1} is 






$$
\mathcal{L}= \frac{1}{2}M\left(\dot{\mathbf x}^2 +\ell^2 \dot \theta ^2 + L^2 \dot \phi^2 + 2 \ell \dot \theta\dot{\mathbf x}\cdot \hat{\mathbf {v}}_\phi\right) - \frac{1}{2}Mg\ell \theta^2,
$$
where
 $\hat{\mathbf {v}}_\phi =(-\sin \phi, \cos \phi)$ is a unit vector perpendicular to the instantaneous direction of the horizontal bar, and $\phi$ is the angle of the bar with respect to the $x$ axis on the plane of the table.


The two  non-holonomic constraints are zero velocity of the point of contact along the direction of the bar (note that the point of contact is at rest with respect to the bar):
\begin{equation}\label{vinculo1}
\hat{\mathbf {u}}_\phi\cdot \dot{\mathbf x}=0,
\end{equation}
and zero velocity in the direction perpendicular to the bar:
\begin{equation}\label{vinculo2}
\dot{\mathbf x}\cdot \hat{\mathbf {v}}_\phi+D\theta \dot \phi =
0,
\end{equation}
with $\hat{\mathbf {u}}_\phi =(\cos \phi, \sin \phi)$ the unit vector in the direction of the bar.

The constraint equations (\ref{vinculo1}) and (\ref{vinculo2}) are linear and can be written in  matrix form  as $\sum_{j=1}^2a_{i,j}(\mathbf q) \dot{q_j}=0$,  with $i=1,2$,  and $\mathbf{q}$ the coordinates $(\mathbf x,\theta,\phi)$. 
As is standard in the treatment of non-holonomic systems \cite{Bloch} we impose the constraints in the equations of motion through Lagrange multipliers
\begin{equation}
{d\over dt} {\partial \mathcal{L}\over \partial \dot q_i}- {\partial \mathcal{L}\over \partial q_i}= \sum_{j=1}^2 \lambda_j a_{j,i}(\mathbf q). 
\end{equation}
Our constrained equations have the form:
\begin{subequations}\label{lorenz}
	\begin{align}
		ML^2\ddot \phi &= -M\ell \dot \theta \dot{\mathbf x}\cdot \hat{\mathbf {u}}_\phi+\lambda _2D \theta   \label{l1} \\
		M\left(\ell^2\ddot \theta + \ell{d\over dt}\dot{\mathbf x}\cdot \hat{\mathbf {v}}_\phi  \right)&= -Mg\ell \theta  \\
		M\left( \ddot {\mathbf x} +\ell{d\over dt}\dot \theta \hat{\mathbf {v}}_\phi \right) &= \lambda_2\hat{\mathbf {v}}_\phi +  \lambda_1\hat{\mathbf {u}}_\phi \label{l33}
	\end{align}
\end{subequations}
From equation (\ref{l33}) and using the constraints we obtain
\begin{equation}\label{lambda2}
\lambda_2=M\left(- D(\dot \phi \dot\theta + \theta \ddot \phi) + \ell \ddot \theta \right).
\end{equation}
Replacing the value of the multiplier in Equation (\ref{l1})  we obtain
\begin{equation}
{
ML^2 \ddot \phi = \lambda_2 D \theta 
\simeq MD\theta \left(- D\dot \phi \dot\theta + \ell \ddot \theta \right),
}
\end{equation}
where we neglected a term $\sim\theta ^2 \ddot \phi$ over $\sim\ddot \phi$, an approximation which we will also adopt in what follows. 

The equations of motion for $\theta $ and $\phi $ become:
\begin{subequations}\label{Aff}
{
\begin{align}
 \ddot \phi &= {D\ell \over L^2} \theta  \ddot \theta - \left({D\over L}\right)^2 \theta \dot \theta \dot \phi 
 \label{phiAf},
 \\
 \ddot \theta &= - {g\over \ell }\theta +\left({D\over \ell}\right) (\theta \ddot \phi +\dot \theta \dot \phi)
 \label{thetBf}.
\end{align}
}
\end{subequations}
Finally, note that from Equation
 (\ref{thetBf}) we have
$$
\theta \dot \theta \dot \phi = {\ell \over D} \ddot \theta \theta + {g\over D}\theta^2 - \theta^2 \ddot \phi,
$$
that replaced in Equation (\ref{phiAf}) leads to the final form of the equations of motion, and constitues one of the results of the present paper:

\begin{subequations}
{
\begin{align}
 \ddot \phi &= -{gD\ell \over L^2} \theta ^2,  \label{FINAL1}
 \\
 \ddot \theta &= - {g\over \ell }\theta +{D\over \ell} \dot \theta \dot \phi.
 \label{FINAL2}
\end{align}
}
\label{FINAL3}
\end{subequations}

\begin{figure}
\includegraphics[scale=0.5]{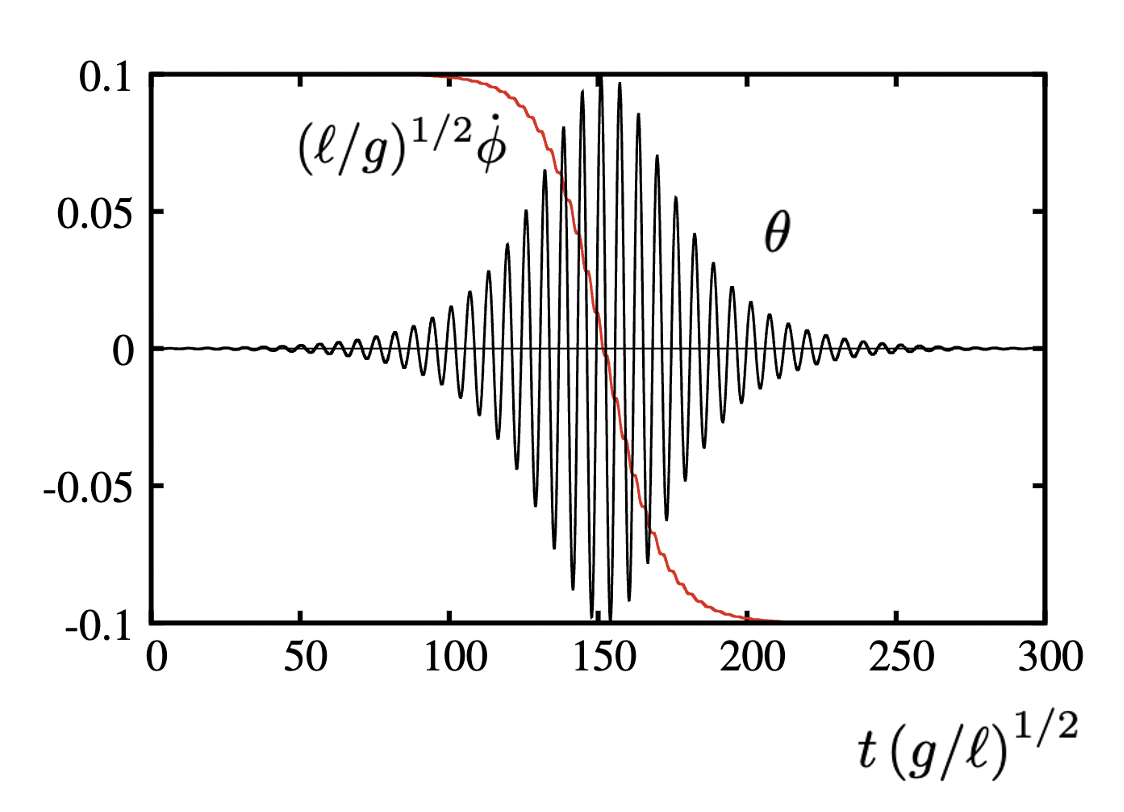}
\caption{Numerical solution  of the single mode rattleback of Equations (\ref{FINAL3}) illustrating the spin inversion accompanied by an increase and decrease of the amplitude of oscillating mode. Initial condition has $\dot\phi=0.1\sqrt{g/\ell}$, and $\theta$ very small. Other parameters used were $\ell/L=1$, $D/\ell=1$.}
\label{numerical}
\end{figure}

The above equations contain the essential elements of spin inversion. Equation (\ref{FINAL2})  describes a harmonic oscillator of amplitude $\theta(t)$ with a friction term with effective friction coefficient $-{\frac{D}{\ell}}  \dot \phi$. This frictional term comprises the back action of 
the $\phi$ mode over $\theta$. Equation (\ref{FINAL1})  corresponds to a torque around the $z$ axis of constant sign, 
in agreement with the qualitative argument presented in Section \ref{Qualitative}. If we start with  $\dot \phi >0$ and $\theta $ infinitesimal, the ``negative friction" term in Equation    (\ref{FINAL2}) gives rise to an increase in amplitude of $
\theta$ and, from equation (\ref{FINAL1}), a simultaneous decrease in the value of $\dot \phi$. This decrease is monotonous as it is proportional to $-\theta ^2$.  When $\dot \phi$ changes sign the corresponding frictional term gives rise to an attenuation of the amplitude of $\theta$ until $\dot \phi$ is constant and negative. This behavior is illustrated in Figure \ref{numerical} where we show a numerical solution of Equations (\ref{FINAL3})


\subsection{The single mode rattleback and the Chaplygin sleigh}

There is a remarkable formal analogy between the present single-mode rattleback and one of the prototypical  non-holonomic mechanical systems: the 
Chaplygin sleigh \cite{Chaplygin}. 
The analogy emerges when considering the previous equations of the single-mode rattleback in terms of slightly different variables. Let us first define
$$
A_\phi = \dot \phi.
$$
We now separate the motion in the periodic mode as $\theta(t) = A_\theta(t) e^{i\omega t}$. We are interested in a situation in which the ``bare" frequency $\omega$ of the mode is large, that is,
$$
\dot \phi \ll \omega= \sqrt{g\over \ell},
$$
and where $A(t)$ is slowly varying in the time-scale of $1/\omega$, that is $\dot A\ll \omega$.
In this regime we are safe to make the following approximation for $\theta ^2$ (replacing $\cos ^2 \omega t$ by it's mean value $\left<|\cos^2 {\omega t}|\right>\sim 1/2$),
$$
\theta ^2 \simeq  {1\over 2} A^2_\theta(t) 
$$
and we are safe to neglect the terms indicated below for
 the  time derivatives of $\theta(t)$:
\begin{eqnarray}
\dot \theta &=&\left\{ \dot A_\theta + i\omega  A  \right\}e^{i\omega t} \simeq  i\omega  A_\theta  e^{i\omega t}, 
\\
\ddot \theta &=&\left\{ \ddot A +2 i\omega \dot A_\theta  - A_\theta\omega^2 \right\}e^{i\omega t} \simeq  \left\{ 2 i\omega \dot A_\theta  - A_\theta\omega^2 \right\}e^{i\omega t}.
\label{SecondDeriv}
\end{eqnarray}

Replacing Equation (\ref{SecondDeriv}) in Equation (\ref{FINAL2}) we obtain,
\begin{eqnarray}
 \left\{ 2 i\omega \dot A_\theta  - A_\theta\omega^2 \right\}e^{i\omega t}  &=& -  A_\theta\omega^2 e^{i\omega t} +{D\over \ell} i\omega A_\theta  e^{i\omega t} \dot \phi
\\
 \dot A_\theta   &=& {D\over 2\ell} A_\theta  A_\phi
\end{eqnarray}

Therefore, we arrive at

\begin{subequations}
{
\begin{align}
 \dot A_ \phi &= -{gD\ell \over 2 L^2} A_\theta ^2  \label{Chap1},
 \\
 \dot A_\theta   &= {D\over 2\ell} A_\theta  A_\phi.
 \label{Chap2}
\end{align}
\label{Chap3}
}
\end{subequations}
These equations have the same structure as those of the Chaplygin sleigh provided we identify the variable $v$ (velocity along the sleigh) with the amplitude 
$\dot \phi= A_ \phi$, and the orientation  of the sleigh $\theta$ with the amplitude of the single oscillatory mode $A_\theta$ of the rattleback.
In fact, the well known tendency of the sleigh to convert its rotational energy into a positive value of $v$ 
corresponds to the property of the single mode rattleback to harvest the kinetic energy of the oscillation, and transform it into
rotational motion around $z$, with a well defined chirality. This remarkable analogy provides a new interpretation of the process of spin inversion in the rattleback, as it shares a close formal analogy with the irreversible dynamics of the Chaplygin sleigh. The physical difference rests in the fact that $v$ in the sleigh corresponds to a linear velocity whereas $A_\phi$ corresponds to an angular velocity. 

\begin{figure}
    \centering
\begin{tikzpicture}
 %
%
\draw[->] (-3.,0)--(3.,0) node [below] {$A_\theta$};
\draw[->] (0.,-4.5)--(0.,4.5) node [above] {$\dot \phi \equiv A_\phi$};
%
%
\draw[black] (0,-4) circle (1.8pt);
\filldraw[white] (0,-3) circle (1.8pt);
\filldraw[black] (0,-2) circle (1.8pt);
\filldraw[black] (0,-1) circle (1.8pt);

\draw [thick] (0,-1) arc
    [
        start angle=-90,
        end angle=90,
        x radius=.5cm,
        y radius =1cm
    ] ;
    \draw [thick,->] (0,-1) arc
    [
        start angle=-90,
        end angle=-30,
        x radius=.5cm,
        y radius =1cm
    ] ;
      \draw [thick,->] (0,-1) arc
    [
        start angle=-90,
        end angle=30,
       x radius=.5cm,
        y radius =1cm
    ] ;
\draw [thick] (0,-2) arc
    [
        start angle=-90,
        end angle=90,
        x radius=1cm,
        y radius =2cm
    ] ;
    \draw [thick,->] (0,-2) arc
    [
        start angle=-90,
        end angle=-30,
        x radius=1cm,
        y radius =2cm
    ] ;
      \draw [thick,->] (0,-2) arc
    [
        start angle=-90,
        end angle=30,
        x radius=1cm,
        y radius =2cm
    ] ;

\draw [thick] (0,-3) arc
    [
        start angle=-90,
        end angle=90,
        x radius=1.5cm,
        y radius =3cm
    ] ;
    \draw [thick,->] (0,-3) arc
    [
        start angle=-90,
        end angle=-30,
      x radius=1.5cm,
        y radius =3cm
    ] ;
      \draw [thick,->] (0,-3) arc
    [
        start angle=-90,
        end angle=30,
    x radius=1.5cm,
        y radius =3cm
    ] ;
\draw [thick] (0,-4) arc
    [
        start angle=-90,
        end angle=90,
        x radius=2cm,
        y radius =4cm
    ] ;
    \draw [thick,->] (0,-4) arc
    [
        start angle=-90,
        end angle=-30,
        x radius=2cm,
        y radius =4cm
    ] ;
      \draw [thick,->] (0,-4) arc
    [
        start angle=-90,
        end angle=30,
     x radius=2cm,
        y radius =4cm
    ] ;
\draw [thick] (0,-1) arc
    [
        start angle=-90,
        end angle=-270,
        x radius=.5cm,
        y radius =1cm
    ] ;
    \draw [thick,->] (0,-1) arc
    [
        start angle=-90,
        end angle=-150,
        x radius=.5cm,
        y radius =1cm
    ] ;
      \draw [thick,->] (0,-1) arc
    [
        start angle=-90,
        end angle=-210,
       x radius=.5cm,
        y radius =1cm
    ] ;
\draw [thick] (0,-2) arc
    [
        start angle=-90,
        end angle=-270,
        x radius=1cm,
        y radius =2cm
    ] ;
    \draw [thick,->] (0,-2) arc
    [
        start angle=-90,
        end angle=-150,
        x radius=1cm,
        y radius =2cm
    ] ;
      \draw [thick,->] (0,-2) arc
    [
        start angle=-90,
        end angle=-210,
        x radius=1cm,
        y radius =2cm
    ] ;

\draw [thick] (0,-3) arc
    [
        start angle=-90,
        end angle=-270,
        x radius=1.5cm,
        y radius =3cm
    ] ;
    \draw [thick,->] (0,-3) arc
    [
        start angle=-90,
        end angle=-150,
      x radius=1.5cm,
        y radius =3cm
    ] ;
      \draw [thick,->] (0,-3) arc
    [
        start angle=-90,
        end angle=-210,
    x radius=1.5cm,
        y radius =3cm
    ] ;
\draw [thick] (0,-4) arc
    [
        start angle=-90,
        end angle=-270,
        x radius=2cm,
        y radius =4cm
    ] ;
    \draw [thick,->] (0,-4) arc
    [
        start angle=-90,
        end angle=-150,
        x radius=2cm,
        y radius =4cm
    ] ;
      \draw [thick,->] (0,-4) arc
    [
        start angle=-90,
        end angle=-210,
     x radius=2cm,
        y radius =4cm
    ] ;

\filldraw[white] (0,-3) circle (1.8pt);
\draw[black] (0,-3) circle (1.8pt);
\filldraw[white] (0,-2) circle (1.8pt);
\draw[black] (0,-2) circle (1.8pt);
\filldraw[white] (0,-1) circle (1.8pt);
\draw[black] (0,-1) circle (1.8pt);
\filldraw[white] (0,-4) circle (1.8pt);
\draw[black] (0,-4) circle (1.8pt);
\filldraw[white] (0,3) circle (1.8pt);
\draw[black] (0,3) circle (1.8pt);
\filldraw[white] (0,2) circle (1.8pt);
\draw[black] (0,2) circle (1.8pt);
\filldraw[white] (0,1) circle (1.8pt);
\draw[black] (0,1) circle (1.8pt);
\filldraw[white] (0,4) circle (1.8pt);
\draw[black] (0,4) circle (1.8pt);
  \end{tikzpicture}
  \caption{Phase portrait of the motion amplitudes of the single mode rattleback.}
    \label{Elipses}
\end{figure}
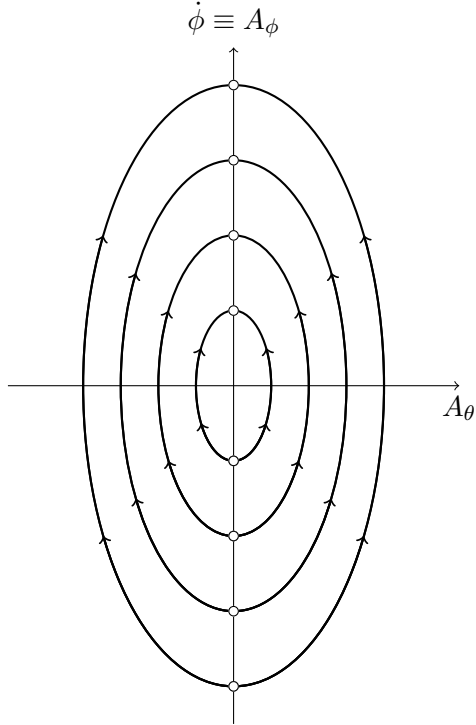

As with the sleigh \cite{Bloch}, 
 Equations (\ref{Chap2}) have a family of equilibria (i.e., points at which the right-hand side vanishes) given by $(A_\theta=0, A_\phi \neq 0)$.
Linearizing about any of these equilibria one finds a zero eigenvalue together with a negative eigenvalue if $A_\phi > 0$ (the stable case) and a positive eigenvalue if $A_\phi < 0$ (the unstable case). 
The solution curves are ellipses in the  $A_\theta, A_\phi$ plane as shown in Figure \ref{Elipses}.

\begin{figure}
\includegraphics[scale=0.45]{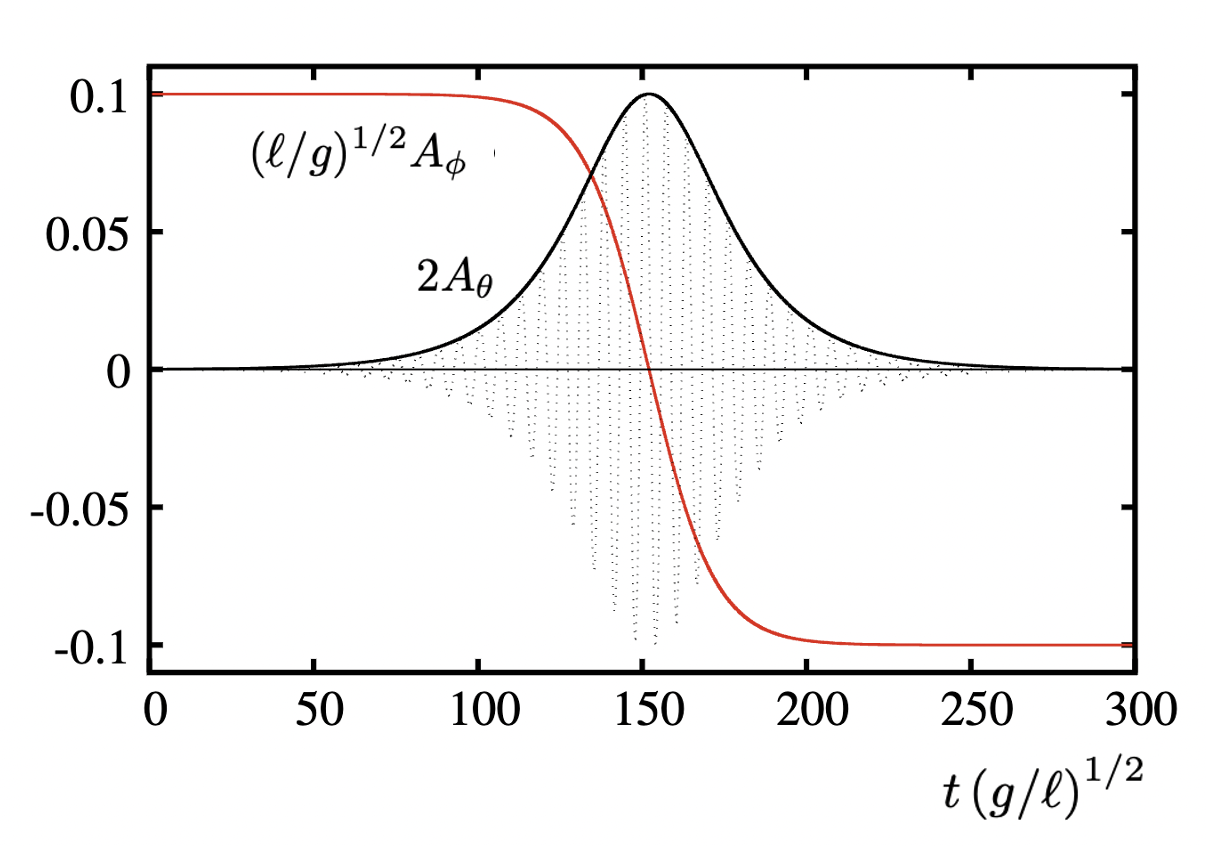}
\caption{Analytical solutions  (Eqs. \ref{an}) of the amplitude  Equations (\ref{Chap3}), for $A_0=0.1\sqrt{g/l}$, $\ell/L=1$, $D/\ell=1$. In dotted lines we also plot
the full solution for $\theta$ (Fig. \ref{numerical}) for comparison.}
\label{amplitudes}
\end{figure}

The time dependence of $A_\theta$, $A_\phi$ can be fully worked out, the final expressions are

\begin{subequations}
\label{an}
\begin{align}
A_\phi&=A_0\tanh\left ({\frac{A_0Dt}{2\ell}}\right)\label{an1}\\    
A_\theta&=\frac{A_0L}{\ell\sqrt{g}}\, {\rm sech}\left(\frac{A_0Dt}{2\ell}\right),
\label{an2}
\end{align}
\label{OneModeExact}
\end{subequations}
where $A_0$ is the asymptotic value ($t\to\pm\infty)$ of $A_\phi$.
In Figure \ref{amplitudes} we show the agreement between these solutions and the the full solution in Figure \ref{numerical}.

\section{Two modes knife edge rattleback}

\label{Twomodes}

The single mode rattleback we discussed in the previous section sheds light on the origin of the coupling mechanism between oscillation and (chiral) rotation. Yet it was presented with a ``prescription" for the shift of the contact point. It is important 
to check if this mechanism, or a similar one, can be  implemented in a well defined mechanical system that includes both the pitching and rolling modes. 
We show here that a fully consistent two-mode rattleback can be constructed starting from the ideas of the previous section.

\begin{figure}
\centering
\includegraphics[width=8cm]{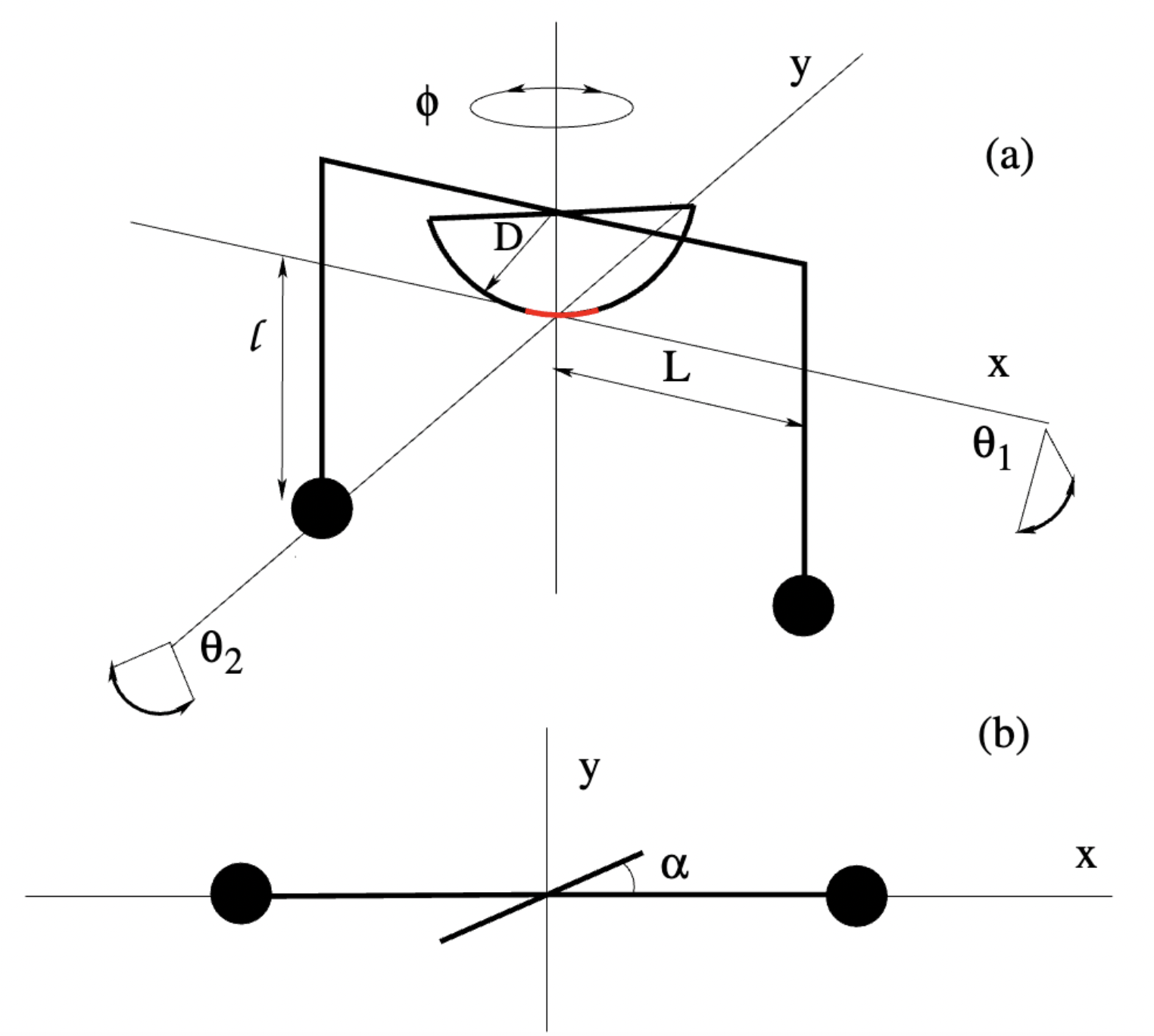}
\caption{(a) Sketch of the knife edge, two--mode rattleback. The disk, bars, and masses define a single rigid body. The basic oscillation ($\theta_1$, $\theta_2$) and rotation $\phi$ modes are indicated. The red 
part of the disk qualitatively indicates the possible contact points depending of the values of $\theta_1$ and $\theta_2$.
(b) Upper view, indicating the definition of the chiral angle $\alpha$.}
\label{sketch_2modes}
\end{figure}

We use a similar geometry of two masses hanging from the ends of a bar with an inverted ``U" form. However, the horizontal part of the bar is modified to set the contact point as sketched in Fig. \ref{sketch_2modes}.
The central portion of the bar has a semi-circular profile of radius $D$. 
The plane of the circle is perpendicular to the horizontal plane and forms an angle
$\alpha$ with the bar, as indicated in Fig. \ref{sketch_2modes}(b).

\begin{figure}
\includegraphics[scale=0.45]{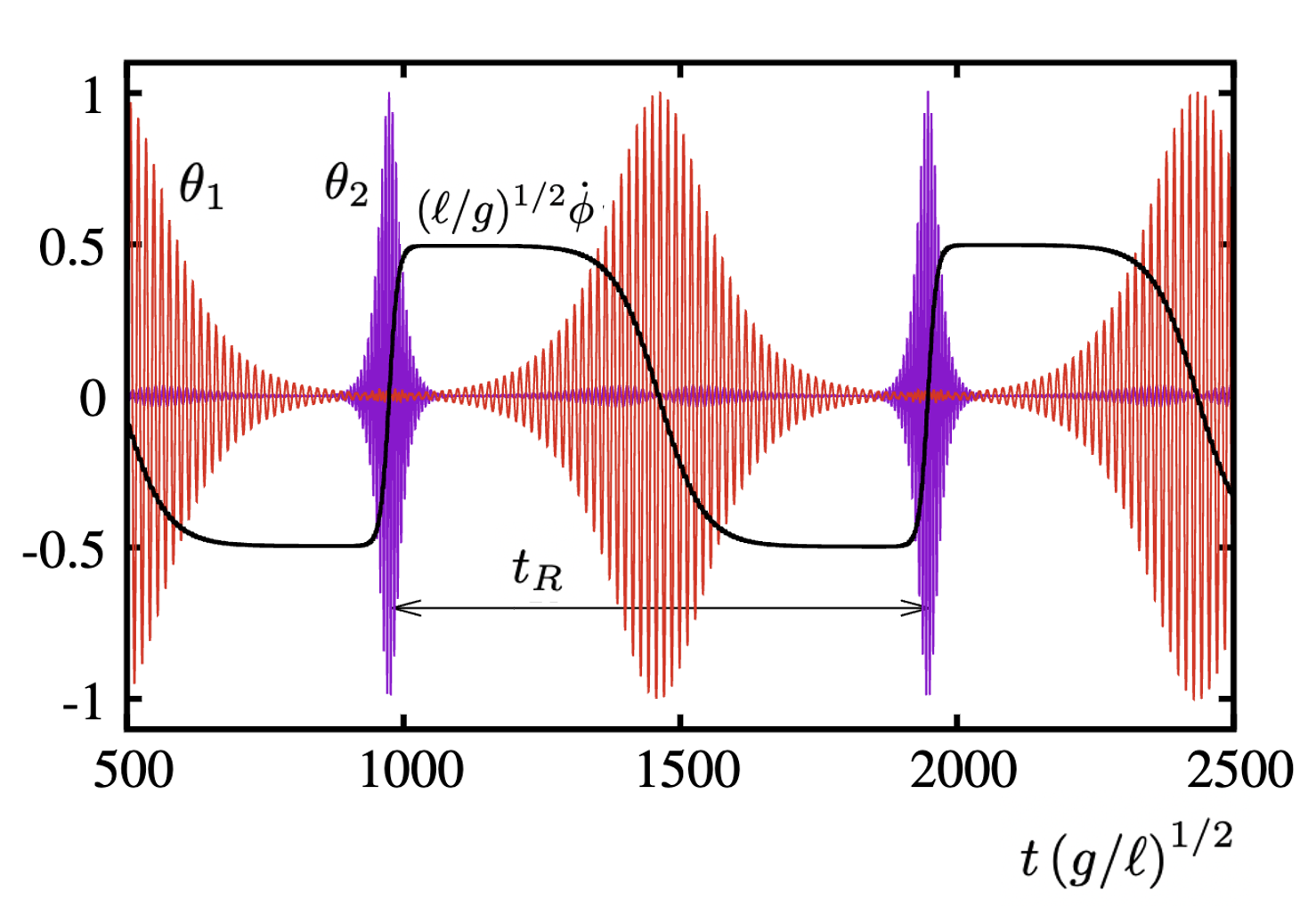}
\caption{Numerical solutions  of Eqs. \ref{eqs_2modes} with  parameters $\ell/L=1/2$, $D/\ell=1/2$, $\alpha=\pi/4$.
Note the periodic reversal of $\dot\phi$, driven alternatively 
by the activation of the $\theta_1$ and $\theta_2$ modes. The reversal time $t_R$ is indicated.}
\label{Numerical2}
\end{figure}

The configuration of the system is determined by two (small) oscillation angles $\theta_1$ and $\theta_2$, the rotation 
angle $\phi$ around $\hat z$, and the position ${\bf x}$ of the middle point of the knife edge. The unconstrained Lagrangian of the system can thus be written as

$$
\mathcal{L}= {1\over 2}M\left\{\dot{\mathbf x}^2 +(L^2+ \ell^2) \dot \theta_2 ^2 +  \ell^2 \dot \theta_1 ^2+ L^2 \dot \phi^2 + 2 \ell \dot{\mathbf x}\cdot\left(\dot \theta_1\hat{\mathbf {v}}
_\phi
-\dot \theta_2 \hat{\mathbf {u}}_\phi\right)
\right\} - {1\over 2}Mg\ell (\theta_1^2+\theta_2^2)
$$
The contact point with the supporting surface is the instantaneous lowest point of the circle, and the constraint is that this physical point must has zero velocity. This leads to the following constraints:

\begin{subequations}
{
\begin{align}
 \dot {\mathbf{x}} \cdot \hat{\mathbf {u}}_\alpha &= 0  \label{Const22}
 \\
 D\theta_\alpha\dot\phi+\dot {\mathbf{x}} \cdot \hat{\mathbf {v}}_\alpha  &= 0 \label{Const23}
\end{align}}
\label{const}
\end{subequations}
where we have defined $\theta_\alpha=(\theta_2\cos \alpha -\theta_1 \sin \alpha)$, and   $\hat{\mathbf {u}}_\alpha$, $\hat{\mathbf {v}}_\alpha$ are horizontal unitary vectors along and perpendicular to the plane of the circle.
From here, the dynamical equations follow

\begin{subequations}
	\begin{align}
		ML^2\ddot \phi &= -M \ell \dot{\mathbf x}\cdot\left(\dot \theta_1\hat{\mathbf {u}}
_\phi
+\dot \theta_2 \hat{\mathbf {v}}_\phi\right)
+\lambda _2 D\theta_\alpha. \label{eqmot1} \\
		M\left(\ell^2\ddot \theta_1 + \ell{d\over dt}\dot{\mathbf x}\cdot \hat{\mathbf {v}}_\phi  \right)&= -Mg\ell \theta_1  \\
		M\left((\ell^2 +L^2)\ddot \theta_2 -\ell{d\over dt}\dot{\mathbf x}\cdot \hat{\mathbf {u}}_\phi  \right)&= -Mg\ell \theta_2  \\
		M\left( \ddot {\mathbf x} +\ell{d\over dt}\left(\dot \theta_1 \hat{\mathbf {v}}_\phi  -\dot \theta_2 \hat{\mathbf {u}}_\phi  \right)\right) &= \lambda_2\hat{\mathbf {v}}_\alpha+  \lambda_1\hat{\mathbf {u}}_\alpha \label{l3}
	\end{align}
\end{subequations}
%
These equations, along with the constraints (Eqs. \ref{const}), enable us to derive the equations of motion in a simplified form by neglecting small terms, as we did previously.
%
%
%
%
%
\begin{subequations}
\begin{align}
\ell^2 \ddot \theta_1&=
 -g\ell\theta_1 +D\ell\dot \theta_\alpha\dot \phi \cos \alpha, 
\label{Const52}\\
(\ell^2+L^2 )\ddot \theta_2&=-g\ell\theta_2 +D\ell\dot \theta_\alpha\dot \phi \sin\alpha\\
L^2\ddot \phi &=-{D^2}\theta_\alpha \dot \theta_\alpha \dot \phi + {\ell D} \theta_\alpha\left(\ddot \theta_1 \cos \alpha
+\ddot \theta_2 \sin \alpha\right) 
\end{align}
\label{eqs_2modes}
\end{subequations}
These equations are the generalization of those of the previous section for the single mode model. Note the similarity in the structure. 
Variables $\theta_1$ and $\theta_2$ are oscillation modes that get an effective friction term proportional to $\dot\phi$, which 
can be positive or negative.
In turn, the variable $\phi$ gets an acceleration that depends quadratically on the $\theta$ variables. Note also that there are terms that are proportional to the product $\theta_1\theta_2$ originated in the non-zero chiral angle $\alpha$.
A qualitatively similar set of equations has been derived by Tokieda and collaborators \cite{Moffatt, Yoshida} using a heuristic approach for the boat-shaped rattleback.
We show in Fig. \ref{Numerical2} a numerical solution of Eqs. \ref{eqs_2modes} with an initial condition having a finite value of $\dot\phi$, and infinitesimal values of $\theta_1$ and/or $\theta_2$ 
(to avoid remaining at an unstable fixed point). From this initial condition the model evolves by periodically reverting the sign of $\dot\phi$ by 
coupling it alternatively to the oscillation modes. In this way, compared to the single mode system, there is not any more a systematic tendency 
to rotate in a single direction, but an alternation between rotation in both senses. In the long run, the average 
value of $\dot\phi$ is zero.

\subsection{Time between spin reversals}

\begin{figure}
\includegraphics[scale=0.45]{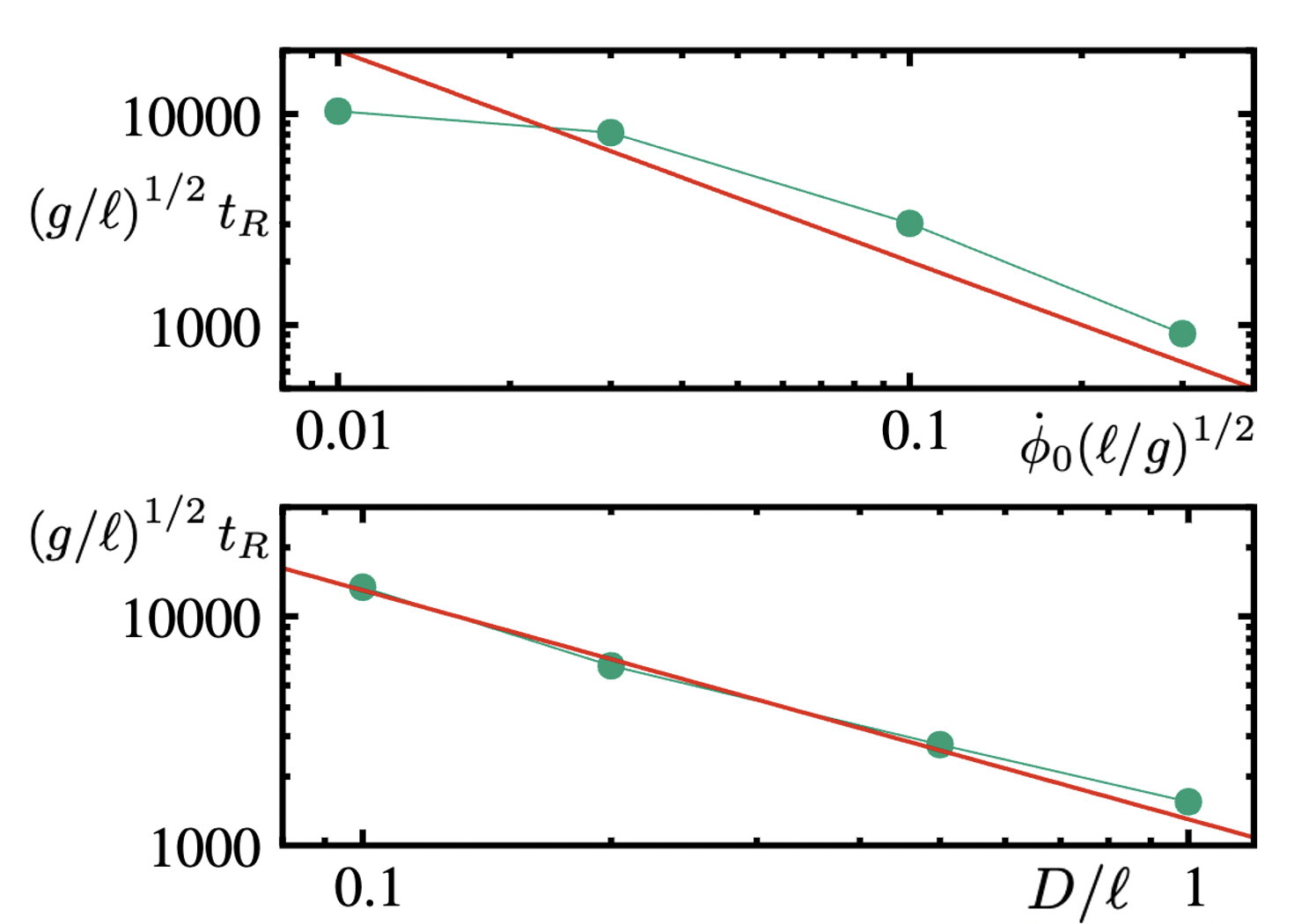}
\caption{Numerical results for the inversion time $t_R$ (see Fig. \ref{Numerical2}) as a function of $\dot\phi_0$ for $D/\ell=1/2$ (a), and as a function of $D/\ell$ for $\dot\phi_0=0.1\sqrt{g/\ell}$(b). In both cases $\ell/L=1/2$. The straight lines depict a power law with exponent $-1$, that is in both cases the expected result for Eq. \ref{tr}. }
\label{tinversion}
\end{figure}

A natural question in the rattleback dynamics concerns the time elapsed between spin reversals. A qualitative understanding can be gained from equations (\ref {FINAL3}) and (\ref{OneModeExact}). Note from  Equation (\ref {FINAL3}) that the the initial conditions $\theta(0)=\dot\theta(0)=0 $ and $\dot \phi(0)= \dot \phi_0$ constitute an unstable situation for which $\theta(t)=0$ and
$\dot \phi(t)=\dot \phi_0$. In order for the spin reversal to take place we need an initial condition $\theta(0) \neq 0 $. 
Now, from equation (\ref{an2}) we see that the time $t_0$ for $\theta $ to increase from a small value $\theta_0\ll 1$ to its maximum $\theta _{\rm max}$
is given by
\begin{equation}
t_0 \simeq {1\over \dot \phi_0} {\ell \over D}  \ln\left({\theta_0\over \theta_{\rm max}}\right).
\label{t0}    
\end{equation}
 In fact
$t_R$ will be $\sim t_0$, if we consider that $\theta_0$ in Eq. \ref{t0} is the background value of the inactive mode, at the maximum amplitude of the active mode driving the inversion.  
Essentially, this is to say that
\begin{equation}
t_R \simeq {1\over \dot \phi_0} {\ell \over D}
\label{tr}
\end{equation}
up to a factor that depends weakly (logarithmically) on the parameters of the model and initial conditions chosen. 
Equation (\ref{tr}) is in agreement with equation (46') in Garcia and Hubbard's treatment \cite{Hubbard2} as well as with Kondo and Nakanishi's paper \cite{Kondo}
We can use this expression to estimate the reversal time $t_R$ in the two mode case, as defined in Fig. \ref{Numerical2}.
We have done
a few simulations to check this expression. In Figure \ref{tinversion} we show the numerically determined value of $t_R$ for different 
parameters, showing an overall good agreement to expression \ref{tr}.

\section{Conclusions}

We presented the chiral knife edge, a new model for a rattleback and showed a full treatment of the model using qualitative arguments, and analytical as well as numerical solution of the non--holonomic equations. We first concentrated on a reduced, one--mode problem which contains the essence of the physics of spin inversion. In short,  a harmonic oscillation $\theta(t)$ requires a restoring force $f\sim \ddot \theta(t)\sim -\theta(t)$. Now, the crucial ingredient of the rattleback is the shift of the contact point from its average positions,  by an amount $\sim\theta$.  Therefore the  restoring force generates a torque around $\hat z$ of value $\sim \theta^2(t)$, that drives spinning in a well defined direction.
In addition we presented a novel --and to us unexpected-- connection between the single mode knife edge and the Chaplygin sleigh, a prototypical non--holonomic system.  We also presented numerical results for the two mode knife edge that illustrate spin inversion in both directions. Finally we presented a qualitative treatment of the time between inversions that agrees with previous results in the literature.
We think the paper offers a new insight on the dynamics of the rattleback, and we plan to explore further consequences in a forthcoming work.
\section{Acknowledgments}

We thank Anthony Bloch for useful comments on the manuscript.

\end{document}